\documentclass[aps,prl,twocolumn,A4paper,groupedaddress,preprintnumbers,showpacs, twoside, floatfix, nofootinbib,amsfonts]{revtex4-1}
\usepackage{amsmath}
\usepackage{color}      
\usepackage{subfigure} 
\usepackage{hyperref}
\usepackage{graphicx}
\newcommand{\lb}{{\langle}}
\newcommand{\rb}{{\rangle}}

\newcommand{\beq}{\begin{equation}}
\newcommand{\eeq}{\end{equation}}
\newcommand{\nn}{\nonumber\\}

\newcommand{\bea}{\begin{eqnarray}}
\newcommand{\ea}{\end{eqnarray}}
\begin{document} 
\begin{abstract} 
We studied the process of decoherence in acoustic black holes. We focused on the ion trap model proposed by Horstmann \emph{et al.} (Phys. Rev. Lett. 104, 250403 (2010)) but the formalism is general to any experimental implementation. For that particular setup, we computed the decoherence time for the experimental parameters that they proposed. We found that a quantum to classical transition occurs during the measurement and we proposed improved parameters to avoid such a feature. We also studied the entanglement between the Hawking-pair phonons for an acoustic black hole while in contact with a reservoir, through the quantum correlations, showing that they remain strongly correlated for small enough times and temperatures.
\end{abstract} 
\title{Decoherence and Loss of Entanglement in Acoustic Black Holes}
\author{Fernando C. Lombardo}
\author{Gustavo J. Turiaci}
\affiliation{Departamento de F\'isica Juan Jos\'e Giambiagi, 
FCEyN UBA, Facultad de Ciencias Exactas y Naturales, 
Ciudad Universitaria, Pabell\'on I, 1428 Buenos Aires, Argentina - IFIBA}
\maketitle
 One of the main results of quantum field theory in curved space-time is the Hawking effect, i.e. the particle creation process that gives rise to a thermal spectrum of radiation outgoing from a black hole \cite{hawking}. This effect together with its entropy complete the interpretation of black holes as thermal objects. On one hand, it is believed that the heart of a theory that unifies quantum mechanics and gravity lies in understanding the nature of this thermality. On the other hand, it is also important to collect experimental evidence in order to gain insight into this phenomenon, but this is practically impossible since black holes' temperatures are less than $nK$. To circumvent this obstacle, W. G. Unruh proposed an analogue gravity hydrodynamical model where phonons propagate in a fluid with a subsonic and supersonic regime \cite{unruh-prl}. This system obeys the dynamics of a massless scalar field near a black hole and provides an experimental implementation to study the Hawking effect. Following this analogy, there have been several proposals that involved BEC \cite{bec}, moving dielectrics \cite{dielectric}, waveguides \cite{nonlinear}, slow light systems \cite{not-slow}, among others. 

 The current proposals do not provide conclusive evidence of the Hawking effect (see for example \cite{quilombo}). Nevertheless, we believe that a particular one by Horstmann \emph{et al.}, \cite{cirac}, provides a promising setup. This system consists of a circular ring of trapped ions moving with an inhomogeneous velocity profile emulating a black hole. The signature of this quantum radiation is the correlation between entangled phonons near the horizon, \cite{corrbec}, which can be measured by coupling the ions' motional degrees of freedom to their internal state, \cite{medicion}. 

Given that we are interested in the quantum nature of the effect, in this letter we study how decoherence affects the measurement, which has not been considered previously. Moreover, we work with the field-theoretical description in order to present a derivation applicable to any implementation of an acoustic black hole. Nevertheless, given our interest in the ion trap we specialize our results to that particular setup.

\emph{The Model}. We can describe the field associated with phonons in $1+1$ dimensions, $\Psi(x,t)$, by means of an action written in the suggestive form
\beq
S_{\mathcal{S}}[\Psi,\partial \Psi]=\iint dtdx \sqrt{-g}g^{\mu\nu}\partial_{\mu}\Psi\partial_{\nu}\Psi,  \label{ac}
\eeq
where the effective metric is of a Painlev{\'e}-Gullstrand-Lema{\^\i}tre type, given by
\begin{equation}
ds^2=(c^2-v^2)dt^2+2vdxdt-dx^2,\label{pgl}
\eeq
and $v(x,t)$ is the local velocity of freely-falling frames. In the ion ring implementation, the ions have the classical trajectories $\theta_i^0(t)$ and the quantum field describes small deviations from them, and it is defined by $\Psi(x=\theta_i^0(t),t)=\sqrt{\rho}\delta\theta_i(t)$, where $\rho$ is the mass density. We take the velocity profile as
\beq
v=\begin{cases}
       v_{\rm min} & 0 \leq \theta \leq \theta_H - \gamma_1\\
	\beta+\alpha\big(\frac{\theta-\theta_H}{\gamma_1}\big)	 & -\gamma_1 \leq (\theta-\theta_H) \leq  \gamma_1\\
       v_{\rm max} & \theta_H+\gamma_1 \leq \theta \leq 2\pi-\theta_H - \gamma_2\\
	\beta-\alpha\big(\frac{\theta-2\pi+\theta_H}{\gamma_2}\big) &-\gamma_2 \leq (\theta-2\pi+\theta_H) \leq \gamma_2\\
       v_{\rm min} & 2\pi-\theta_H+\gamma_2 \leq \theta \leq 2\pi
      \end{cases} \nn
\eeq
where $\beta=(v_{\rm max}+v_{\rm min})/2$ and $\alpha=(v_{\rm max}-v_{\rm min})/2$. The minimum and maximum velocities are related by the condition that each ion has to make one revolution during the period $\tau$. It is important to notice that we use an approximate velocity profile. The actual one must be smoother, but ours is a good approximation for our calculations. Taking this into account, the parameters $\gamma_i$ and $\theta_H$ do not exactly match those of \cite{cirac}. 

Although the problem can be treated in a discrete fashion, we choose to stick to the field description since the action given in Eq.(\ref{ac}) is common to every implementation of acoustic black holes, not restricted to ion traps. Then our procedure can be used to study the decoherence in any acoustic black hole.

In this letter we propose to work out the non-equilibrium features of the interaction between the ion trap and its environment, since these are responsible for decoherence. As usual for this kind of tasks, we use the Schwinger-Keldysh or \emph{in-in} formalism. Our system is described by the action given in Eq.(\ref{ac}). Following the quantum Brownian motion (QBM) paradigm \cite{calhu}, the environment is described by a continuous array of bosonic harmonic oscillators, distributed in each position of the circular trap, and represented by the degree of freedom $\hat{q}_{\nu}(\theta,t)$ with action
\beq
S_{\mathcal{E}}[q_{\nu}]=\frac{1}{2}\int_0^{\infty} d\nu I(\nu) \int d^2x   \big[ \dot{q}^2_{\nu}(\theta,t)-\nu^2 q_{\nu}^2(\theta,t)\big].
\eeq
The function $I(\nu)$ is the mass of each environmental oscillator. The interaction between the system and the environment is given by the action
\beq
S_{\rm int}[\Psi,q_{\nu}]=-\tilde{\gamma} \int_0^{\infty} d\nu\int d^2x~\Psi(x) q_{\nu}(x),
\eeq
so that the total action is given by $S[\Psi,q_{\nu}]=S_{\mathcal{S}}[\Psi]+S_{\mathcal{E}}[q_{\nu}]+S_{\rm int}[\Psi,q_{\nu}]$. For example, in the ion trap proposed in \cite{cirac}, the velocity profile is produced by an external electric field, which is generated with surface-electrodes. Irregularities over their surface produce fluctuations and the heat bath models each mode of the fluctuating field. Details of the possible nature of the environment is extensively discussed in \cite{wineland}. The fact that we are thinking of a fluctuating field coupled with the coordinate of the ions justifies the bilinear coupling used here. To simplify the treatment we also assume an ohmic bath, although more general environment would not change the results \cite{entorno}. The bath is at rest with respect to the laboratory. Moreover, the reservoir naturally has the same discretization (Planckian) length-scale as the ions in the ring, and therefore it is also sub-Planckian. Certain effects developed in the presence of a trans-Planckian environment, such as Miles-type instability \cite{miles}, are absent here. 

The elements of the density matrix are given by $\rho(\Psi^+,q^+;\Psi^-,q^-|t)=\lb\Psi^+,q^+|\widehat\rho(t)|\Psi^-,q^-\rb$. We assume an initial uncorrelated thermal state between system-environment, both at temperature $T=\hbar(k_B\beta)^{-1}$. More general initial states would not change substantially the process of decoherence, see \cite{estadoinicial}.The reduced density matrix, $\rho_{\rm r}(\Psi^{\pm}|t)$, is defined as usual, performing a trace over the environment. 

We studied the evolution of the system following standard procedure, such as \cite{lombardo}. The evolution of the reduced density matrix is derived from the effective action $S_{\rm eff}[\Psi^+,\Psi^-]=S_{\mathcal{S}}[\Psi^+]-S_{\mathcal{S}}[\Psi^-]+S_{\text{IF}}[\Psi^+,\Psi^-]$, where the index $\pm$ designates both branches of the time path, see \cite{calhu}, and the Feynman-Vernon influence action is given by
\beq
S_{\text{IF}}=\int d^2xd^2x'\Delta(x) \bigg({\bf D}(x,x')\Sigma(x')+\frac{i}{4}{\bf N}(x,x')\Delta(x')\bigg), \nn
\eeq
where $\Delta(x)=\Psi^+(x)-\Psi^-(x)$ and $\Sigma=\frac{1}{2}(\Psi^+(x)+\Psi^-(x))$. Both the noise and dissipation kernels are local in space. Their exact expressions are the QBM ones,
\bea
{\bf D}(t,&t'&)=\int_0^{\infty}d\nu\frac{\tilde{\gamma}^2}{I(\nu)\nu}\sin{\nu(t-t')}\Theta(t-t'),  \\
{\bf N}(t,&t'&)=\int_0^{\infty}d\nu\frac{\tilde{\gamma}^2}{I(\nu)\nu}\coth{\frac{\beta\nu}{2}}\cos{\nu(t-t')}, 
\ea
where $\tilde{\gamma}^2/\nu I(\nu)=\gamma^2 \nu$ for an ohmic bath.
From $S_{\rm eff}$, we can write the semiclassical Langevin equation of the field, given by
\beq
\frac{1}{\sqrt{g}}\partial_{\mu}\big(\sqrt{g}g^{\mu\nu} \partial_{\nu}\Psi \big)+\int ds{\bf D}(t,s)\Psi(s,x)=\xi(x,t). \label{eq:lang}
\eeq
The field $\xi$ is a stochastic force with a gaussian probability density with zero mean and correlation $\lb\xi(x)\xi(x')\rb=\hbar {\bf N}(x,x')$. 

\emph{Decoherence Time}. The problem with this approach is that the velocity profile depends on time in an awkward fashion during the collapse, making the master equation hard to derive. To avoid this drawback we first studied the imaginary part of the effective action $\text{Im}(S)[\Psi_{\rm cl}]$, verifying numerically that the small period, named $\tau_c$, when the velocity changes with time does not contribute to the decoherence time. We also show that the small angular region where the velocity is not constant is also irrelevant. Using these results, we will derive the master equation using the weak coupling approximation. 

For piecewise space and time independent velocity profiles, the modes can be found using the usual null coordinates
\bea
{\sf u}&=& t-\int \frac{dx}{c(x)+v(x)}=t-x_u,\\
{\sf v}&=& t+\int \frac{dx}{c(x)-v(x)}=t-x_v.
\ea
 Each solution can be decomposed in $u$ and $v$ modes which have the form $\Psi_{u/v}(x,t)=\Psi \cos{(\omega {\sf u}/{\sf v})}$. For fixed $\omega$, we studied the decoherence of each mode $u$ and $v$, but we found that the result does not depend on this choice. Since the spatial dimension is compact we find the available $\omega$ demanding that $\Psi_{{\sf u}/{\sf v}}(t,x+L)=\Psi_{{\sf u}/{\sf v}}(t,x)$. This gives a numerable set of frequencies, and the maximum frequency is found by estimating that the wavelength cannot be smaller than the ions' separation, which we call $\delta \equiv L/N$, with $L$ the circumference of the trap. The steps towards the derivation of the master equation are standard procedure, which can be found in \cite{lombardo,calhu}. The diffusion coefficient of the master equation is $d(t)=\int_0^t ds \cos(\omega s) {\bf N}(t,t-s)$ and we call $V=\int_0^L dx \cos^2{(\omega x_{{\sf u}/{\sf v}})}$. The solution to the master equation is approximately $\rho_r \propto  \exp \{-\Gamma \int_0^t dt d(t)\}$ where $\Gamma= \gamma^2 V (\Psi^+-\Psi^-)^2/2$. The condition to find the decoherence time is $\Gamma \int_0^{t_D} dt \ d(t) \approx 1$.

Since the proposal consists of measuring the entanglement between phonons next to the event horizon, we study the decoherence of two modes with the best resolution we could have to obtain an upper bound for $t_D$, taking $(\Psi^+-\Psi^-)\approx \sqrt{\rho}\delta$. At $T=0$, the diffusion coefficient is, for an ohmic bath, $d(t)\approx\omega\text{Si}(\omega t)$ \cite{paula}. We also assume that we are able to see several oscillations of the mode before it decoheres, i.e. $t\gg \omega^{-1}$. Then $d(t)\approx \omega \pi/2$ and $t_D(T=0)\approx2\hbar\gamma^{-2}\delta^{-2}(\omega \pi V)^{-1}$.

In order to compute the decoherence time, we need an estimation for $\gamma$. In \cite{cirac}, an upper bound for the stochastic force $\xi$ in units of the mean force on the ions $\overline{F}$, $\xi=\zeta \overline{F}$, was established, given by $\zeta \leq \zeta_{\rm max}=5\times 10^{-6}$, see Eqs. (80) and (81) of the second paper in \cite{cirac}. We relate $\zeta$ with $\gamma$ through the r.m.s. value of the stochastic force present in Eq. (\ref{eq:lang}), using the $\xi$'s two point function. The relation between these variables is $\gamma=\zeta\sqrt{2\rho/\hbar}(v_{\text{max}}-v_{\text{min}})$.
\begin{figure}[h!]
\begin{center}
\subfigure{\includegraphics[scale=0.34]{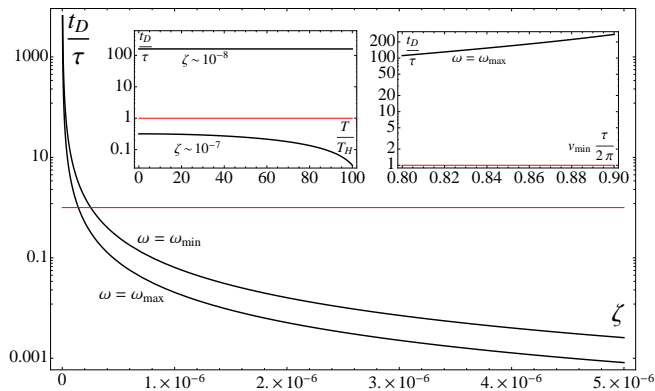}}
\end{center}
\caption{\label{tdec}Plot of the decoherence time as a function of $\zeta$, for maximum and minimum frequency, the velocity (right inset) and the system's initial temperature (left inset).}
\end{figure}

Fig.\ref{tdec} shows how the decoherence time depends on $\zeta$. First, we check that $\omega_{\rm max} t_D \sim 10^5 \gg1$ and the approximations used are consistent with the results. We note that for the stochastic force bound proposed in Ref.\cite{cirac}, decoherence is too strong, since the measurement lasts no more than $\tau$ seconds due to classical instability issues of the system, so $t_D$ must be bigger than $\tau$ in order to be able to measure quantum Hawking radiation. Let's say that the decoherence time must be a couple of orders of magnitude larger, $t_D \sim 100 \tau$, then $\zeta$ must be less than $\zeta_{\rm max}=2\times10^{-8}$. We also show the dependence of the decoherence time on the velocity profile in Fig.\ref{tdec}. We plot $t_D$ as a function of $v_{\rm min}$, for $\zeta_{\rm max}$, concluding that it does not change substantially since the dependence is smooth. Of course, for low enough velocities $t_D$ can be less than $\tau$, since it is monotone. Nevertheless, $v_{\rm min}$ must be close to $(2\pi/\tau) 0.8\overline{3}$ in order to produce an event horizon.

For the case of small temperatures, we separate the $\nu$ integral between two intervals $(0,2\beta^{-1})$ such that for small enough frequencies $\beta\nu/2\ll1$, $\coth{(\beta\nu/2)}\approx 2/\beta\nu$, and for high enough frequencies $(2\beta^{-1},\infty)$ such that $\beta \nu /2 \gg 1$, one can take $\coth{(\beta\nu/2)} \approx 1$. For times longer than the period of the mode, $\omega t \gg 1$, the diffusion coefficient has the lower bound $\int_0^{t_D} dt \ d(t,\beta)=\omega \pi t_D/2+4\omega^{-2}\beta^{-2}$, resulting in a decoherence time of 
\begin{equation}
t_D(T)\approx \frac{\hbar^2}{\zeta^2 (v_{\rm max}-v_{\rm min})\delta^2\omega\pi\rho^2V}-\frac{4}{\omega^3\pi\beta^2}+\mathcal{O}(\beta^{-3}).
\end{equation}
In the Fig.\ref{tdec} it is shown the dependence of the decoherence time with temperature for several couplings. For $\zeta \sim 10^{-8}$, the decoherence time is much larger than $\tau$ for a wide range of temperatures. Therefore, it is clear that our proposed coupling is a significant improvement from the one proposed in \cite{cirac}, and composes the most adequate parameters for the desired experimental conditions. In \cite{cirac} it is proven that it takes some time to generate the entanglement between the Hawking phonons, even in an equilibrium situation, and that the entanglement is only appreciable for temperatures below $100T_H$. Taking this into account, and even though it takes a large temperature to affect the quantum features of the system with such a small coupling as $\zeta_{\rm max}$, our results are only useful to measure the Hawking effect below $100T_H$. Moreover, it can be verified that the low temperature limit is only valid below $\sim 100T_H$.

\emph{Correlations}. Hawking radiation can be understood as a pair production of virtual particles, one of which falls into the black hole and the other, outgoing, becomes real, building up the Hawking radiation \cite{hawking}. The role of the correlations between this pair of particles has been studied as a signature of the quantum Hawking effect, for example \cite{corrbec}-\cite{unruh-cor}. It was found that the entanglement between this pair is translated to a sharp peak of $\langle \Pi^-(x_1,t)\Pi^-(x,t) \rangle$ as a function of $x$, where $\Pi$ is the canonical momentum conjugate to $\Psi$ and $\Pi^-$ corresponds to its left-moving modes, $x_1$ is inside the black hole and $x$ is outside. This magnitude was calculated using the Israel-Hartle-Hawking state, \cite{unruh-cor}, and it was also calculated numerically in the case of the circular ion trap, \cite{cirac}. 

We derived this quantum correlation and obtained an analytic expression that depends on time and also on the initial temperature of the system. We also show that before decoherence the difference between the closed and open system correlation is indistinguishable, but for later times the correlations change substantially from the closed case one. To obtain the correlation we follow the procedure presented in \cite{calzetta}, Eq.(4.8), which consists of solving the Langevin equation, multiplying the solutions to construct the correlation and then integrating over the initial state and the stochastic source. Since the system is weakly coupled to its environment we treat the dissipation kernel term perturbatively, calling $\Psi_o$ and $G_{\rm ret}$ the solution of the free wave equation and its Green function; thus the full solution is $\Psi_o+G_{\text{ret}} \cdot (\xi - {\bf D} \cdot \Psi_o)$.
\begin{figure}[h!]
\begin{center}
\includegraphics[scale=0.37]{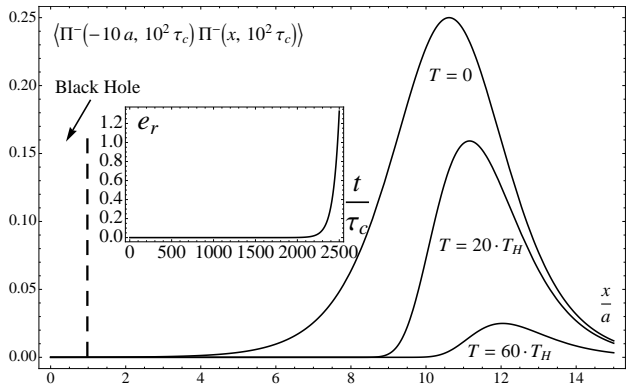}
\end{center}
\caption{\label{pipi} Plot of the momentum-momentum correlation for $x_1=-10a$ and $t=100\tau_c$. The profile parameters are $v_{\rm min}=0.9$, $v_{\rm max}=1.1$. It is shown for different temperatures. Inset: Relative contribution of the correlation due to the environment as a function of time, in units of the collapse time $\tau_c$.}
\end{figure}

We first obtain the analytic expression without environment. For this we first compute the left-moving modes in a background, given by Eq.(\ref{pgl}), which satisfies $\partial_L \Psi^-=0$, where $\partial_L=(\partial_t+v \partial_x-\partial_x)$. The profile is $v(|x|<a,t)=\sigma(t)(1-\kappa x)$, $v(x>a,t)=\sigma(t) v_{\min}$, $v(x<a,t)=\sigma(t) v_{\max}$, where $\sigma(t)$ is a function that satisfies $\sigma(0)=0$ and $\sigma(t\gg\tau_{c})=1$, and $\tau_{c}$ is the time of collapse of the acoustic black hole. We solve the equation using the method of characteristics, well known from fluid dynamics of compressible fluids \cite{fluid}. We normalized the modes in a way that resembles the usual Minkowski expansion at $t=0$. 

When solving by the method of characteristics we get two solutions; for example, we get the curves in the $x>a$ region and the $x<a$ region. The curves originated in $x_0<a$ must match along the boundary $x=a$. These solutions present the Hawking radiation-related peak. Nevertheless, when the characteristic starts from $x=a$ or bigger, then the full solution is the $v(x>a,t)=\sigma(t) v_{\min}$ one. In this case the peak is not present. Therefore, the characteristic with $x_0=a$ provides a boundary between the region where the horizon is important and the region where it is not. For the regime where the Hawking correlation is present, the result is plotted for different temperatures in Fig.\ref{pipi}. As noted in \cite{cirac}, the correlations are present even when the ions' temperature is higher than the Hawking temperature, and this seems to be a general feature of the Hawking effect, as long as decoherence is not present. 

We also estimate the relative influence due to the presence of the environment, $e_r\equiv |(C_c(k)-C_o(k))/C_c(k)|$, where $C_c$ is a single mode of the correlation $\langle\Pi_1\Pi_2\rangle$ when the system is closed and $C_o$ is the correlation when the environment is turned on. In the inset of Fig.\ref{pipi} we plotted $e_r$ as a function of time, for low wavenumbers, $e_r(k\sim0,t)$ since it decreases with increasing $k$. For small times, $e_r \sim 0$ and then the environment does not affect the correlation. For large enough times, the relative correction due to the environment becomes of the same order as the correlation of the closed system. We expect that after this happens, the correlation gets distorted, the peak gets lost and the entanglement between the Hawking pair gets lost due to decoherence.

If an experimental proposal is to be taken seriously, then decoherence must be taken into account. We hope that this study reassures that the measurement of the Hawking effect is possible.
We also provide a derivation of the correlation between the Hawking phonons as a function of time and temperature and we check that the relative contribution of the environment to this magnitude is irrelevant. We hope that this would help to deepen the understanding of the Hawking effect in acoustic black holes, and the entanglement between the Hawking-pair.

\textit{Acknowledgements}. This work was supported by UBA, CONICET and ANPCyT. GJT is also supported by a CIN fellowship. The authors thank Juan Ignacio Cirac and Diana L\'opez Nacir for useful discussions.
\addcontentsline{toc}{section}{References}


\begin{thebibliography}{30} 
\bibitem{hawking}
S. Hawking,
Nature {\bf 248}, 30 (1974); 
Commun. Math. Phys. {\bf 43}, 199 (1975);
J. Hartle \& S. Hawking,
Phys. Rev. D {\bf 13}, 2188 (1976).

\bibitem{unruh-prl}
W.G.~Unruh,
Phys. Rev. Lett. {\bf 46}, 1351 (1981).

\bibitem{bec}
L.J. Garay \textit{et al.},
Phys. Rev. Lett. {\bf 85}, 4643 (2000);
Phys. Rev. A {\bf 63}, 023611 (2001);
L.J.~Garay,
Int. J. Theor. Phys. {\bf 41}, 2073 (2002);
C.~Barcelo \textit{et al.},
Class. Quant. Grav. {\bf 18}, 1137 (2001).

\bibitem{dielectric}
R.~Sch\"utzhold \textit{et al.},
Phys. Rev. Lett. {\bf 88}, 061101 (2002);
I.~Brevik \& G.~Halnes,
Phys. Rev. D {\bf 65}, 024005 (2002).

\bibitem{nonlinear}
W.G.~Unruh \& R.~Sch\"utzhold,
Phys. Rev. Lett. {\bf 95}, 031301 (2005).

\bibitem{not-slow}
W.G.~Unruh \& R.~Sch\"utzhold,
Phys. Rev. D {\bf 68}, 024008 (2003).

\bibitem{quilombo}
F. Belgiorno \emph{et al.}, 
Phys. Rev. Lett. {\bf 105}, 203901(2010);
W.G.~Unruh \& R.~Sch\"utzhold,
\emph{ibid}. {\bf 107}, 149401 (2011);
Belgiorno \emph{et al.},
\emph{ibid}. {\bf 107}, 149402 (2011).

\bibitem{cirac}
B. Horstmann \textit{et al.},
Phys.Rev.Lett. {\bf 104}, 250403 (2010);
New J. Phys., 13, 045008 (2011).

\bibitem{corrbec}
R. Balbinot \textit{et al.},
Phys. Rev. A {\bf 78}, 021603 (2008).

\bibitem{medicion}
C. Monroe \emph{et al.}, 
Phys. Rev. Lett. 75, 4011 (1995);
D.J. Wineland \emph{et al.}
J. Res. NIST 103, 259 (1998);
R. Sch\"utzhold, 
Phys. Rev. Lett. 97, 190405 (2006).

\bibitem{calhu}
E.A. Calzetta \& B.L. Hu,
\emph{Nonequilibrium Quantum Field Theory},
(Cambridge U. Press, Cambridge, 2008).

\bibitem{wineland}
S. Schneider \& G.J. Milburn,
Phys. Rev. A 59, 3766 (1999);
C.J. Wyatt \emph{et al.},
Nature, Vol. 403, 269 (2000);
Seidelin \emph{et al.},
Phys. Rev. Lett. {\bf 96}, 253003 (2006);

\bibitem{entorno}
B.L. Hu \textit{et al.}, 
Phys. Rev. D 47, 1576 (1993); 
J.P. Paz \textit{et al.}, 
\textit{ibid.} 47, 488 (1993); 
F.C. Lombardo \& P.I. Villar, 
Phys. Lett. A 371, 190 (2007).

\bibitem{miles}
W.G. Unruh \& R. Sch\"utzhold,
Phys. Rev. D 71, 024028 (2005).

\bibitem{estadoinicial}
L.D. Romero \& J.P. Paz, Phys. Rev. A 55, 4070 (1997)

\bibitem{lombardo}
F.C. Lombardo \& F.D. Mazzitelli,
Phys. Rev. D {\bf 53}, 2001(1996);
F.C. Lombardo \textit{et al.}, 
Nucl. Phys. B {\bf672}, 462 (2003);
F.C. Lombardo \& D.L\'opez Nacir,
Phys. Rev. D {\bf 72}, 063506 (2005). 

\bibitem{paula}
F.C. Lombardo \& P.I. Villar,
Phys. Lett. A {\bf 336}, 16 (2005).

\bibitem{unruh-cor}
R. Sch\"utzhold \& W.G. Unruh,
Phys. Rev. D {\bf 81}, 124033 (2010).

\bibitem{calzetta}
E.A. Calzetta \textit{et al.},
Phys. A, {\bf 319}, 188-212 (2003).

\bibitem{fluid}
A.H. Shapiro,
{\em The Dynamics and Thermodynamics of Compressible Fluid Flow.},
The Ronald Press Company, New York (1953).
\end{thebibliography}
\end{document}